\documentclass[twocolumn]{pasj01}
\usepackage[utf8]{inputenc}
\usepackage{graphicx}
\usepackage{multirow}
\usepackage{xcolor}
\usepackage{caption}
\usepackage{comment}
\usepackage[margin=22mm]{geometry} 
\usepackage[switch,mathlines]{lineno}
\usepackage{amstext}

\begin{document}

\title{The candidates of long-periodic variable sources in 6.7 GHz methanol masers associated with four high-mass star-forming regions}

\author{Yoshihiro \textsc{tanabe}\altaffilmark{1}}
\author{Yoshinori \textsc{yonekura}\altaffilmark{1}}

\altaffiltext{1}
{Center for Astronomy, Ibaraki University, 2-1-1 Bunkyo, Mito, Ibaraki 310-8512, Japan }

\email{yoshihiro.tanabe.ap@vc.ibaraki.ac.jp}

\maketitle

\begin{abstract}
Results of the long-term monitoring observations of the 6.7 GHz Class II methanol masers associated with the four high-mass star-forming regions by Hitachi 32-m radio telescope are presented.
We detected periodic flux variability in G06.795$-$0.257, G10.472$+$0.027, G12.209$-$0.102, and G13.657$-$0.599 with the periods of  968, 1624, 1272, and 1266 d, respectively, although the detected period is tentative due to the short monitoring term relative to the estimated period.
The facts that the flux variation patterns show the symmetric sine curves and that the luminosities of the central protostar and periods of maser flux variation are consistent with the expected period-luminosity (PL) relation suggest that the mechanisms of maser flux variability of G10.472$+$0.027 and G12.209$-$0.102 can be explained by protostellar pulsation instability.
From the PL relation, central stars of these two sources are expected to be very high-mass protostars with a mass of $\sim$40 $M_{\odot}$ and to have the mass accretion rate of $\sim$2$\times$10$^{-2}$ $M_{\odot}$yr$^{-1}$.
On the other hand, G06.795$-$0.257 and G13.657$-$0.599 have the intermittent variation patterns and have luminosities that are an order of magnitude smaller than those expected from PL relation, suggesting variation mechanisms of these sources originated from binary system.
Since almost all the maser features  vary with the same period regardless of its geometry, periodic accretion model may be appropriate mechanisms for flux variability in G06.795$-$0.257 and G13.657$-$0.599.

\end{abstract}
 
\section{Introduction}
High-mass stars have significant impact on their surrounding environment and evolution of galaxies through several feedback mechanisms, e.g., stellar winds, ultraviolet radiation, and supernovae.
Therefore understanding the processes of the formation and evolution of high-mass stars is one of the most important issues in astronomy.
However, our understanding of the formation processes of high-mass stars remains inadequate, hampered by the observational difficulties that they are born in distant and very deeply embedded in dense gas.
Class II methanol maser is a well established tracer of high-mass star-forming regions (HMSFRs) (e.g, Minier et al.\ 2003; Ellingsen 2006; Pandian et al.\ 2007; Breen et al.\ 2013).
The strongest methanol maser line is 6.7 GHz the $5_{1}-6_{0}A^{+}$ transition, discovered firstly by Menten (1991) and over 1000 6.7 GHz methanol masers have been discovered by unbiased surveys and targeted observations (e.g., Pestalozzi et al.\ 2005; Pandian et al.\ 2007; Caswell et al.\ 2010, 2011; Green et al.\ 2010, 2012; Breen et al.\ 2015; Yang et al.\ 2017, 2019; Rickert et al.\ 2019; Ortiz-Le\'{o}n et al.\ 2021; Nguyen et al.\ 2022).
The 6.7 GHz maser is sensitive to the local physical conditions around high-mass protostars, and thus is an excellent observational probe to study the high-mass star formation process.

Several 6.7 GHz methanol masers have presented their characteristic periodic flux variations.
Goedhart\ et\ al.\ (2003)\ first discovered a periodic flux variation of 243.3 d in the 6.7 GHz methanol masers associated with HMSFR G9.62$+$0.20E, and so far, 31 periodic methanol maser sources have been reported (Tanabe, Yonekura and  MacLeod\ 2023, hereafter TYM23, and references therein, Aberfelds et al.\ 2023).
Their periods range from 23.9 d (for G14.23$-$0.50 reported by Sugiyama et al.\ 2017) to 1265 d (for G5.900$-$0.430 reported by TYM23).
The variation patterns in the time series of the 6.7 GHz methanol maser flux are classified into two types: continuous such as a sinusoidal curve and intermittent with quiescent phase, and differences in variation patterns are thought to be due to differences in the origin of 6.7 GHz maser periodicity.
Such flux variabilities were also detected in  H$_2$O, OH, and H$_2$CO masers (e.g., Szymczak et al.\ 2016; Seidu et al.\ 2022).
Several explanations of mechanism for 6.7 GHz methanol maser periodicity have been proposed;
colliding wind binary (CWB) system (van der Walt\ 2011; van der Walt et al.\ 2016), protostellar pulsation (Inayoshi et al.\ 2013), spiral shock in a circumbinary system (Parfenov \& Sobolev\ 2014), periodic accretion in a circumbinary system (Araya et al.\ 2010), and eclipsing binary system (Maswanganye et al.\ 2015). 
These models can explain the maser flux variations of some sources well, however, there is no clear consensus on the general mechanisms of maser periodicity. 

In this paper, we present new discoveries of periodicities in 6.7 GHz methanol maser sources in four HMSFRs G06.795$-$0.257, G10.472$+$0.027, G12.209$-$0.102, and G13.657$-$0.599, and discuss the mechanisms of their flux variability.

\section{Observations}
Monitoring observations of the 6.7 GHz methanol masers were made with the Hitachi 32m telescope of Ibaraki station, a branch of the Mizusawa VLBI Observatory of the National Astronomical Observatories Japan (NAOJ), operated jointly by Ibaraki University and NAOJ (Yonekura et al. 2016).
This is as a part of the Ibaraki 6.7 GHz Methanol Maser Monitor\ (iMet) program\footnote{http://vlbi.sci.ibaraki.ac.jp/iMet/}.

Monitoring observations began on 2013 Jan.
The cadence of observations is once per every $\sim$10 d from the start of the monitoring observations to 2015 Aug., and once per every $\sim$5 d from 2015 Sep.\ to the present. 
We use the data up to Sep. 30, 2023 in this paper.
Observations after 2014 May were made at about the same azimuth and elevation angle to minimize intensity variations due to systematic telescope pointing errors.
The half-power beam width of the telescope is $\sim$ \timeform{4.6'} with the pointing accuracy better than $\sim$ \timeform{30"} (Yonekura et al.\ 2016). 
The coordinates of target sources adopted in our observations are summarized in table \ref{obstab}.
The reference for these coordinates is Green et al.\ (2010).

Observations are made by using a position-switching method. 
The OFF position is set to $\Delta {\rm R.A.} = +\ \timeform{60'}$ from the target source. 
The integration time per observation is 5 minutes for both the ON and OFF positions.
A single left circular polarization (LCP) signal was sampled at 64 Mbps (16 mega-samples per second with 4 bit sampling) by using a K5/VSSP32 sampler (Kondo et al.\ 2008).
The recorded bandwidth is 8 MHz (RF:\ 6664--6672 MHz) and they are divided into 8192 channels.
After averaging over 3 channels, the 1-sigma root-mean-squares noise level is approximately 0.3 Jy and the velocity resolution is 0.13 km s$^{-1}$.
The antenna temperature was measured by the chopper-wheel method and the system noise temperature toward the zenith including the atmosphere (${T}^{*}_{\rm sys}$) is typically 25--35 K.
In our monitoring program, we observe $\sim$60 methanol maser sources per day, of which the variation of the flux density of sources that do not show the intrinsic variation are less than $\sim$20\%.

\begin{table}[h!]
\centering
\caption{List of source}
  \begin{tabular}{ccc}
  \hline
 Source&R.A.&Dec. \\
  &(J2000)&(J2000)\\\hline
  G06.189-0.358 & 18 01 57.75 & $-$23 12 34.9 \\ 
  G10.472$+$0.027 & 18 08 38.20 & $-$19 51 50.1 \\
  G12.209$-$0.102 & 18 12 39.92 & $-$18 24 17.9 \\
  G13.657$-$0.599 & 18 17 24.26 & $-$17 22 12.5 \\
  \hline
  \end{tabular}
  \label{obstab}
  \begin{tabnote}
  \end{tabnote}
\end{table}

\section{Results}

\subsection{Method of the periodic analysis}
The periodicity was estimated by employing the Lomb-Scargle (LS) periodogram method (Lomb 1976 and Scargle 1982) and the asymmetric power function (PF) fitting.

LS method can give us false detection of the period, and thus we adopted the false alarm probability functions (Frescura, Engelbrecht, and Frank 2008) which is a probability of judging noise as a real signal.
In this paper, if the peak value of the power spectrum is higher than the 0.01\% false alarm level, we decided the obtained period is reliable.
The error of periods obtained by the LS method are estimated as the half width of half maximum (HWHM) of each peak in the periodogram.
It should be noted that the estimated error is large because the observation term is not much longer than the obtained period.

The asymmetric PF is given by the equation modified from Szymczak et al.\ (2011):
\begin{equation}
S(t)=A{\rm exp}\lbrace{s(t)}\rbrace +Ct+D\ ,
\end{equation}
where {\it A}, {\it C}, and {\it D} are constants and 
\begin{equation}
s(t)=\frac{-B\ {\rm cos}(2\pi\frac{t}{P}+\phi)}{1-f\ {\rm sin}(2\pi\frac{t}{P} +\phi)}\ .
\end{equation}
Here {\it B} is the amplitude relative to the average value, {\it P} is the period, $\phi$ is the phase at $t=0$, and $f$ is the asymmetric parameter defined as the rise time from the minimum to the maximum flux, divided by the period and takes the value of $ -1\ <\ f\ <\ 1 $\  (Szymczak et al.\ 2011).
The power function is strictly symmetric when $f=0$, while the peak appears earlier as {\it f} approaches $-$1 and later as {\it f} approaches 1.
In this paper, we defined that the power function is symmetric when $ -0.5\leq f \leq 0.5 $, and asymmetric when $|{\it f}|\ >\ 0.5$.
We also defined the full width between half maximum (FWHM) of the flare.
A variation pattern is defined as intermittent if the estimated FWHM of the flare is less than one third of the detected period.

Variation pattern analysis is complemented by the phase-folded light curves.
Details of the analysis are shown in the Appendix.

\subsection{G06.795$-$0.257}
\begin{figure}[!htb]
 \begin{center}
  \includegraphics[width=8.5cm,bb= 0 0 576 432 ]{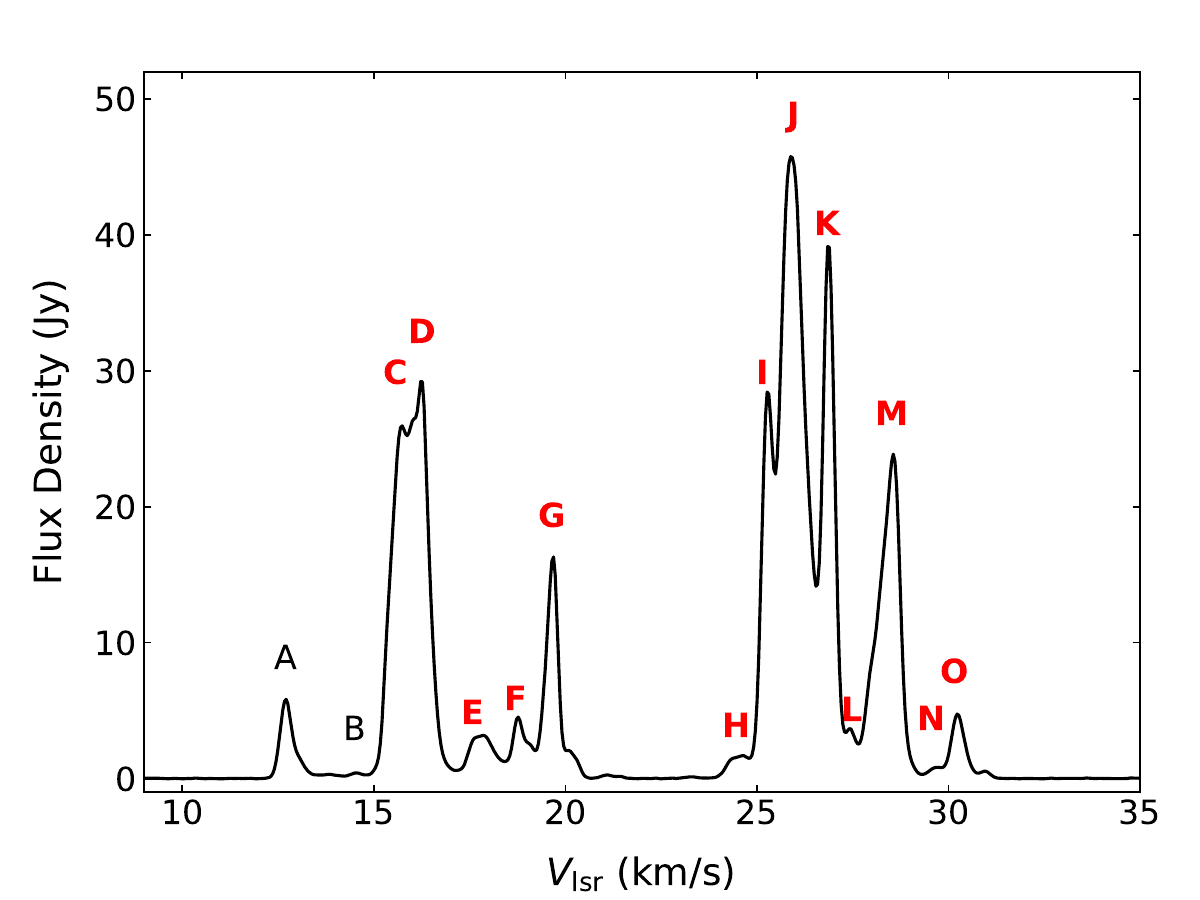}
   \caption{Averaged spectrum of the 6.7 GHz methanol maser associated with G06.795$-$0.257.
   All 580 scans (MJD = 56295--60217) of 5 minutes integration are averaged and 3$\sigma$ detection limit is 0.038 Jy.
    Labels A to O indicate spectral components at $V_{\rm LSR} =$ 12.72, 14.55, 15.76, 16.24, 17.79, 18.78, 19.68, 24.58, 25.29, 25.94 26.87, 27.45, 28.55, 29.57, and 30.26 km s$^{-1}$
   , respectively. 
   In 13 features colored in red (C to O), the periodicity is detected in this paper.
   }
    \label{sp1}
 \end{center}
\end{figure}

\begin{figure*}[Htb!]
 \begin{center}
  \includegraphics[width=17.5cm, bb=0 0 1584 1998]{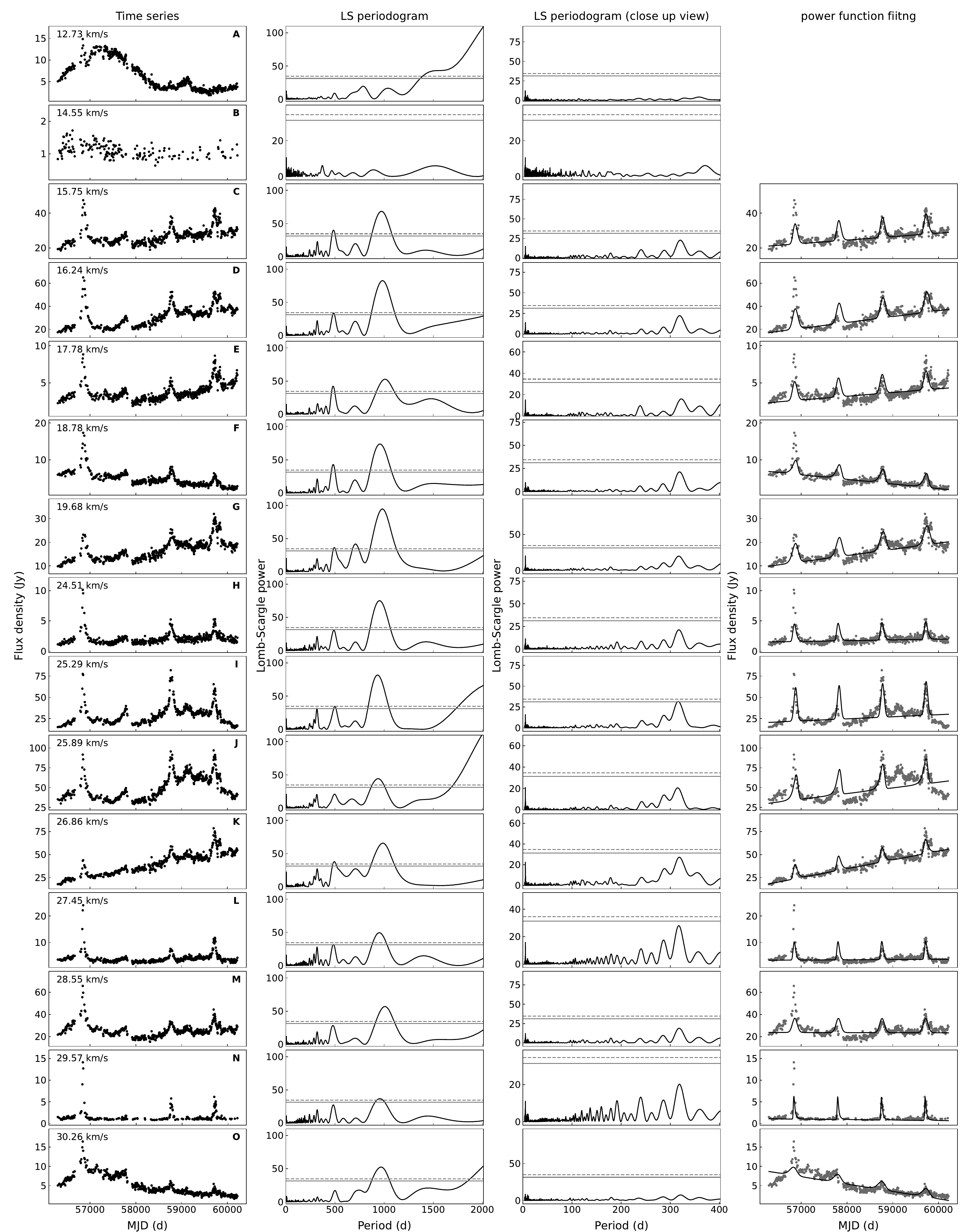}
   \caption{Time series and results of periodic analysis for G06.795$-$0.257.
   The first column show the time series of flux density of each velocity feature.
   We excluded the data points whose flux densities are less than 3 $\sigma$.
   The second and third columns show the Lomb-Scargle power spectra plots and that of close up view, respectively.
   The solid and dotted lines in second and third columns represent the 0.01\% and 0.001\% false alarm probability levels.
   The fourth columns show the best fitting of periodic power function.
   It should be noted that we did not make a fitting for the features which LS method cannot detect the period.
 In the first column, labels of alphabet represent velocity features.}
    \label{ana1}
 \end{center}
\end{figure*}

The averaged spectrum obtained by the integration of all 544 scans of G06.795$-$0.257 is presented in figure \ref{sp1}.
In this source, 15 velocity features are distributed from $V_{\rm LSR}\ \sim$12 km s$^{-1}$ to $\sim$31 km s$^{-1}$.
Among the 15 velocity features in this source, 13 periodic features are detected by our periodic analysis and results are summarized in table \ref{tab1} and figure \ref{ana1}.
The detected periods by LS method are between 931 and 1010 d and those by PF fitting are between 947 and 988 d for each velocity feature.
In the third column of table \ref{tab1}, if multiple reliable peaks were detected, the period with the highest power is shown in bold.
All the detected periods by two methods are consistent within estimated errors.
The flux variation pattern is intermittent for all the velocity features with very small FWHM compared to the estimated periods.
In LS method, the averaged period with the reliable LS power is 968 d.
Harmonic peaks, i.e. P = $\sim$1000/n where n = 2 and 3, are seen in the periodograms for label C to N features.
Asymmetric parameters range from $-$0.69 to 0.97, which indicates that the variability pattern of this source includes an asymmetric with fast rising, asymmetric with fast decaying, and symmetric.

\begin{table*}[!h]
\begin{center}
\centering
\caption{Parameters of the periodic features in G06.795$-$0.257}
  \begin{tabular}{cclcccc}
  \hline
  label&$V_{\rm LSR}$&$ P_{\rm LS} $\footnotemark[$*$]&$P_{\rm PF}$\footnotemark[$\dag$]&$f$\footnotemark[$\ddag$]& FWHM\\
  &(km s$^{-1}$)&(d)&(d)&&(d)\\\hline
  C&15.75 & 969\ (98)& 954\ (2) & 0.34\ (0.20)   &105\\
  D&16.24 & 985\ (104)& 958\ (2) & 0.07\ (0.20)   &123\\
  E&17.78 &   985\ (97) & 958\ (3) &  0.05\ (0.24) &115\\
  F&18.78 &   953\ (95) & 953\ (2) &  $-$0.69\ (0.07) &119\\
  G&19.68 &  723(47),\ \textbf{1002}\footnotemark[$\S$]\ (113)& 956\ (3) & -0.12\ (0.19)   &117\\
  H&24.51 &  946\ (98)& 957\ (2) & 0.76\ (0.10)     &94\\
  I&25.29 &   931\ (128) & 952\ (2) &  $-$0.68\ (0.21) &92\\
  J&25.89 &   953\ (85) & 947\ (3) &  $-$0.22\ (0.18) &93\\  
  K&26.86 &   1010\ (130) & 951\ (2) &  0.69\ (0.18) &95\\
  L&27.45 &   953\ (95)& 957\ (1) & 0.78\ (012)     &70\\
  M&28.55 &   994\ (100)& 954\ (3) & $-$0.01\ (0.28)   &115\\
  N&29.57 &   953\ (101) & 956\ (1) &  0.97\ (0.11)  &42\\
  O&30.26 &   953\ (86) & 988\ (8) &  $-$0.07\ (0.20) &214
  \\\hline
  \end{tabular}
  \label{tab1}
  \end{center}
    \begin{tabnote}
  \footnotemark[$*$]Period estimated by LS method.\\
  \footnotemark[$\dag$]Period estimated by power function fitting.\\
  \footnotemark[$\ddag$]Asymmetric parameter.\\
   \footnotemark[$\S$]The bold represents the highest LS power if multiple reliable periods are detected.
  \end{tabnote}
\end{table*}

\subsection{G10.472$+$0.027}
\begin{figure}[h!]
 \begin{center}
  \includegraphics[width=8.5cm,bb= 0 0 576 432 ]{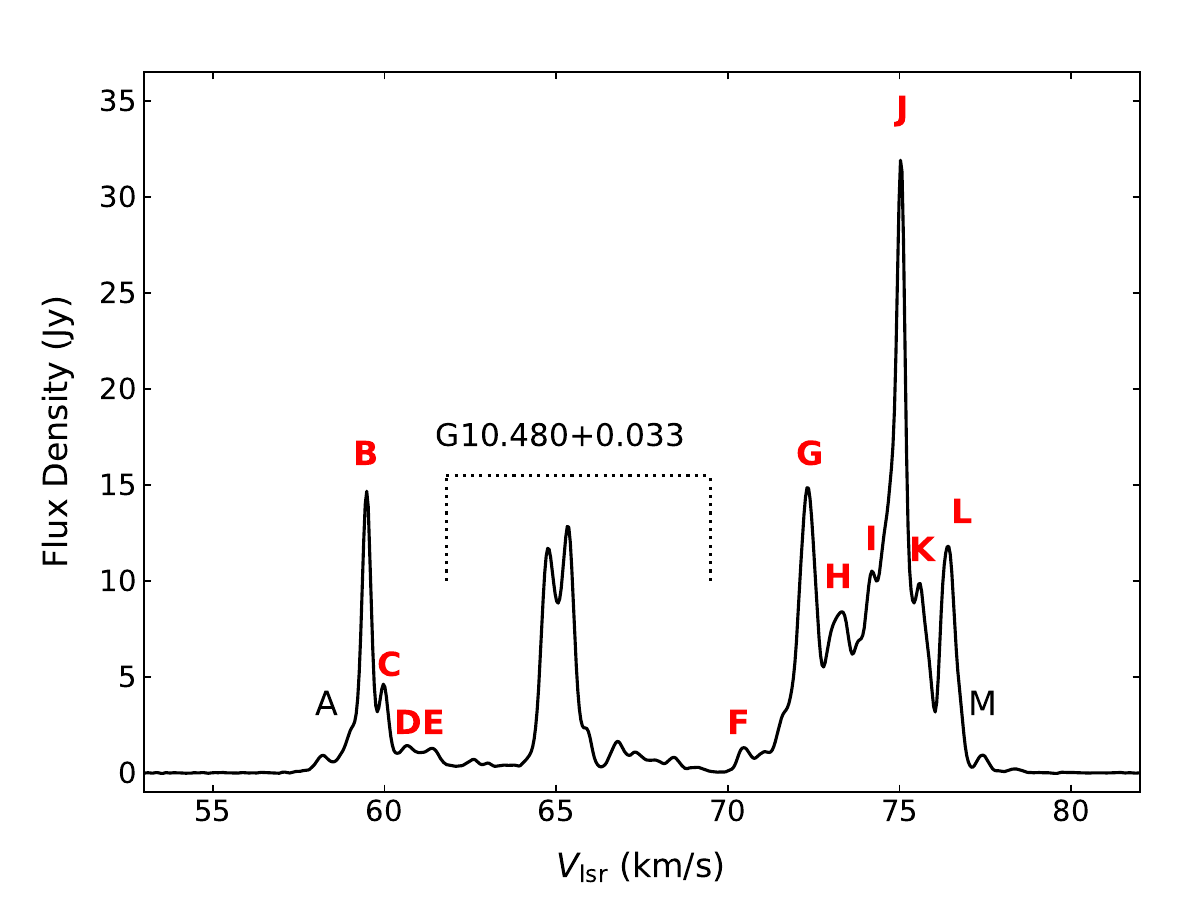}
   \caption{Same as in figure \ref{sp1}, but for G10.472$+$0.027.
   All 576 scans (MJD = 56295--60208) of 5 minutes integration are averaged and 3$\sigma$ detection limit is 0.040 Jy.
    Labels A to M indicate spectral components at $V_{\rm LSR} =$ 58.21, 59.48, 59.97, 60.64, 61.37, 70.47, 72.32, 73.31, 74.21, 75.04, 75.58, 76.41, and 77.38 km s$^{-1}$
   , respectively. 
   In 10 features colored in red (except A, D, and M), the periodicity is detected in this paper.
   }
    \label{sp2}
 \end{center}
\end{figure}

\begin{figure*}[H!tb]
 \begin{center}
  \includegraphics[width=17.5cm, bb=0 0 1584 1732]{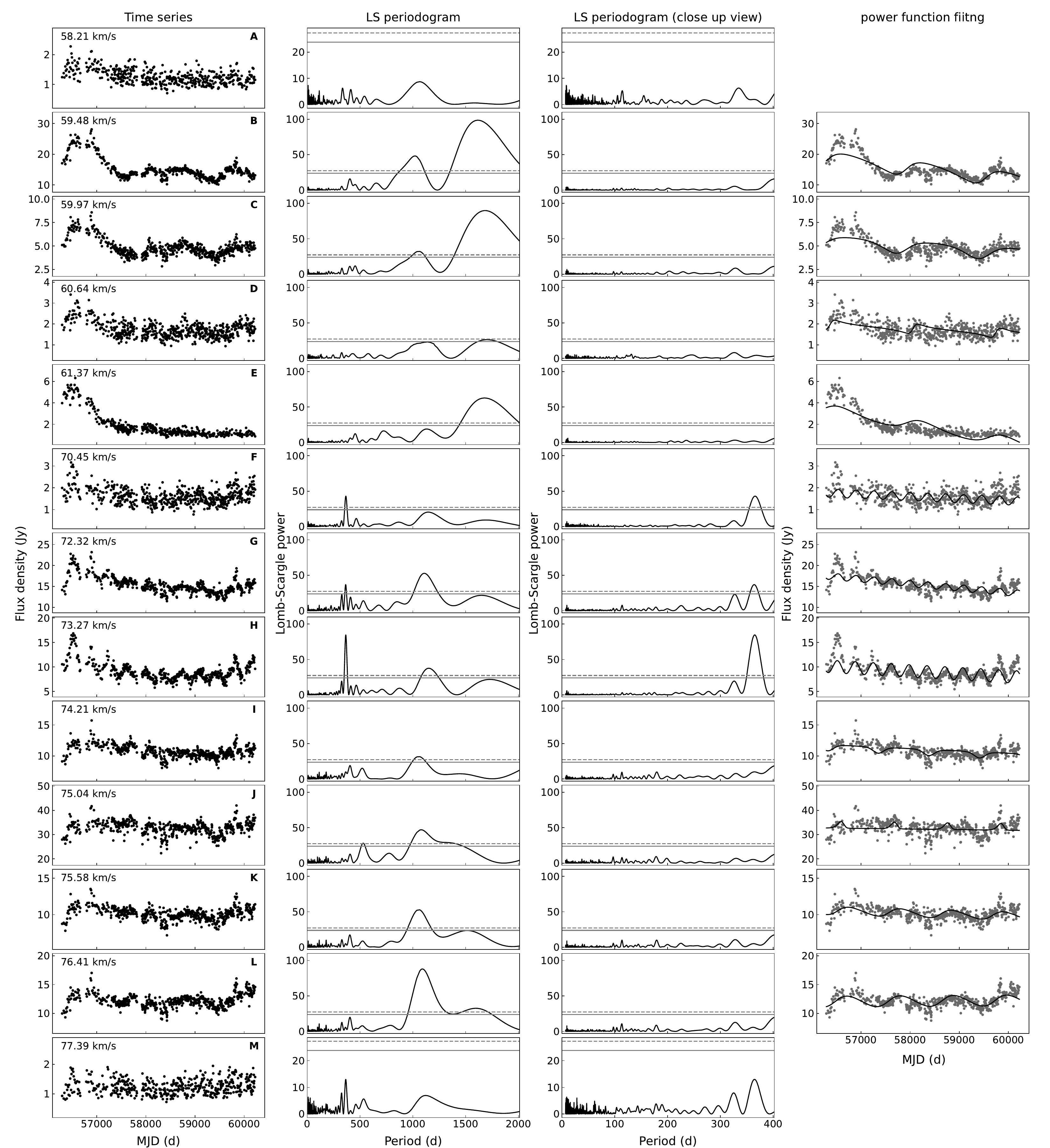}
   \caption{Same as in figure \ref{ana1}, but for G10.472$+$0.027.  
 }
    \label{ana2}
 \end{center}
\end{figure*}

The averaged spectrum obtained by the integration of all 526 scans of G10.472$+$0.027 is presented in figure \ref{sp2}.
According to Green et al.\ (2010), Hu et al.\ (2016), and Nguyen et al.\ (2022), the features between $V_{\rm LSR}=62$ km s$^{-1}$ and $V_{\rm LSR}=70$ km s$^{-1}$ are associated with HMSFR G10.480$+$0.033, approximately 34 arcsec apart from G10.472$+$0.027.
Therefore we exclude these features in this paper and detected 13 velocity features in G10.472$+$0.027.
The LS method detected periodicity in 11 of the 13 features and results of the periodic analysis are summarized in table \ref{tab2} and figure \ref{ana2}.
In the third column of table \ref{tab2}, if multiple reliable peaks were detected, the period with the highest LS power is shown in bold.
We detected period of over 1500 d in the blue-shifted, B, C, D, and E features.
On the other hand, the red-shifted features has a relatively shorter periods compared to the blue-shifted features.
The F, G, and H features have a period of 366 d and the I, J, K, and L features have a period of $\sim$1000 d.
The averaged periods of the these three groups are 1623, 366, and 1020 d, respectively.
All the features experience continuous flux variation and the features with periods longer than 1500 d show asymmetric flare patterns with fast rise and slow decay.
All the velocity features with a period of 366 d (features F, G, and H) have a $f$ of $\sim$0 and FWHM/$P_{\rm LS}$ of $\sim$0.5, i.e., sinusoidal-continuous variation pattern.
Except the feature I, all the velocity features with a period of $\sim$1000d (features J, K, and L), experience a symmetric continuous variation ($f \sim0$ and FWHM/$P_{\rm LS}\sim0.5$).
Flares with a period of 366 d for features F, G, and H are detected 11 times during the monitoring term, thus the detections of this period are considered reliable.

\begin{table*}[!ht]
\begin{center}
\caption{Parameters of the periodic features in G10.472$+$0.027}
  \begin{tabular}{cclcccc}
  \hline
  label&$V_{\rm LSR}$&$ P_{\rm LS}$ &$P_{\rm PF}$& $ f$&FWHM \\
  &(km s$^{-1}$)&(d)&(d)&&(d)\\\hline
  B&59.48 & 979\ (64), \textbf{1604}\ (256) & 1608\ (21) &  $-$0.53\ (0.07) &998\\
  C&60.64 & 889\ (76), 996\ (79), \textbf{1652}\ (305)& 1670\ (26) & $-$0.17\ (0.07)   &1034\\
  D&61.37 & 1698\ (291)& 1684\ (35) & $-$0.66\ (0.15) &982\\
  E&61.37 & 1540\ (274)& 1540\ (31) & $-$0.60\ (0.07) &764\\
  F&70.45 & 366\ (28)& 365\ (14) & $-$0.02\ (0.21) &178\\
  G&72.32 & \textbf{366}\ (24), 540\ (49), 1070\ (209)& 365\ (2) & $-$0.19\ (0.20) &179\\
  H&73.27 & \textbf{366}\ (25), 1090\ (205) & 365\ (1) & $-$0.02\ (0.13) &183\\
  I&74.21 & 996\ (201)& 1021\ (7) & $-$0.89\ (0.07) &877\\
  J&75.04 & 540\ (79), 750\ (92), \textbf{1050} (202)& 1053\ (17) & 0.16\ (0.19) &583\\
  K&75.58 & 1013\ (198)& 1016\ (14) & 0.29\ (0.16) &538\\
  L&76.41 & 1050\ (225)& 1053\ (10) & 0.43\ (0.11) &561
  \\\hline
  \end{tabular}
  \label{tab2}
    \end{center}
\end{table*}

\subsection{G12.209$-$0.102}
\begin{figure}[!htb]
 \begin{center}
  \includegraphics[width=8.5cm,bb= 0 0 576 432 ]{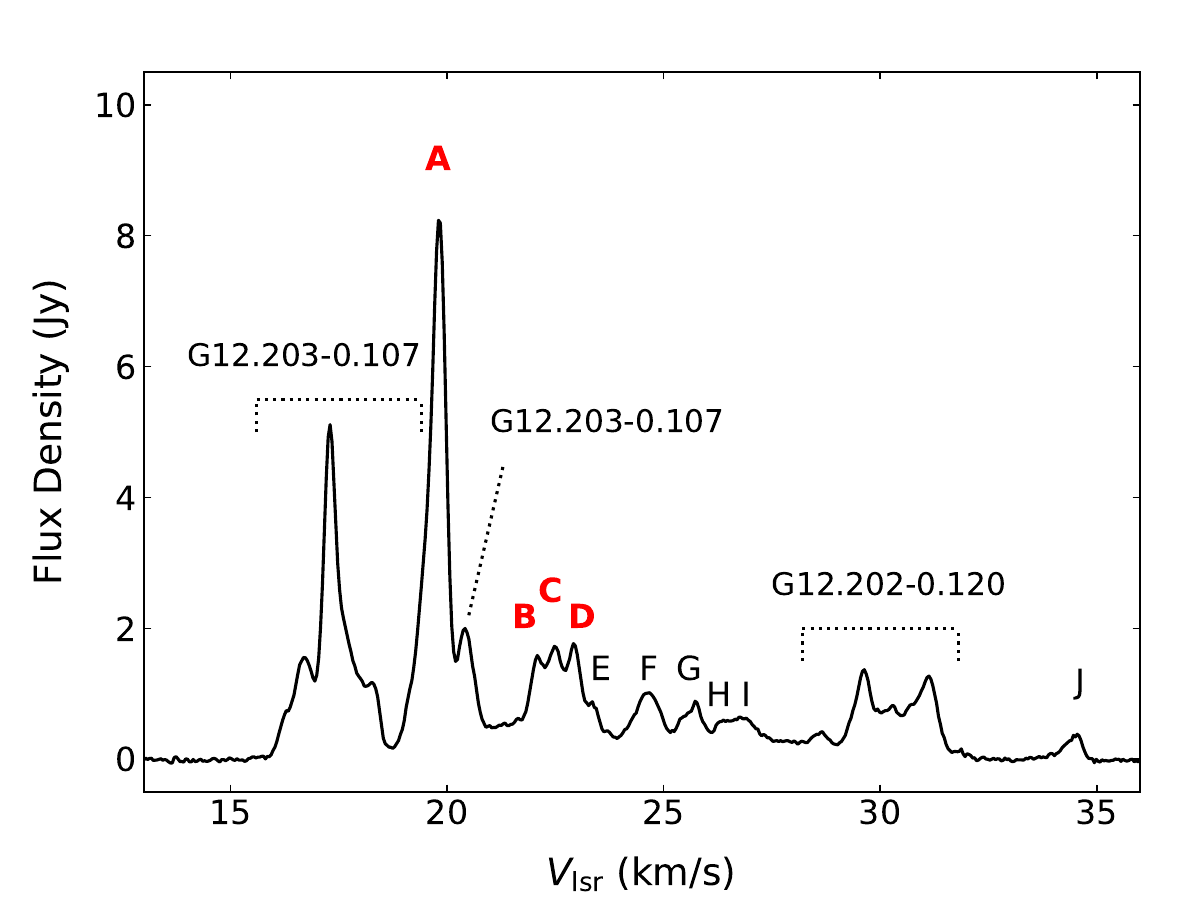}
   \caption{Same as in figure \ref{sp1}, but for G12.209$-$0.102.
   All 561 scans (MJD = 56296--60214) of 5 minutes integration are averaged and 3$\sigma$ detection limit is 0.041 Jy.
   Labels A to J indicate spectral components at $V_{\rm LSR} =$ 19.82, 22.08, 22.44, 22.97, 23.38, 24.63, 25.71, 26.32, 26.78, and 34.52 km s$^{-1}$
   , respectively.
   In four features colored in red (A to D), the periodicity is detected in this paper.}
    \label{sp3}
 \end{center}
\end{figure}

The averaged spectrum obtained by the integration of all 511 scans of G12.209$-$0.102 is presented in figure \ref{sp3}.
There are two other maser sources in our beam, viz. G12.203$-$0.107 and G12.202$-$0.120 (Caswell et al.\ 1995; Hu et al.\ 2016; Nguyen et al.\ 2022).
The velocity ranges of these three source are close to one another, thus some velocity components may be blended with G12.209$-$0.102.

\begin{figure*}[h!tb]
 \begin{center}
  \includegraphics[width=17.5cm, bb=0 0 1584 1332]{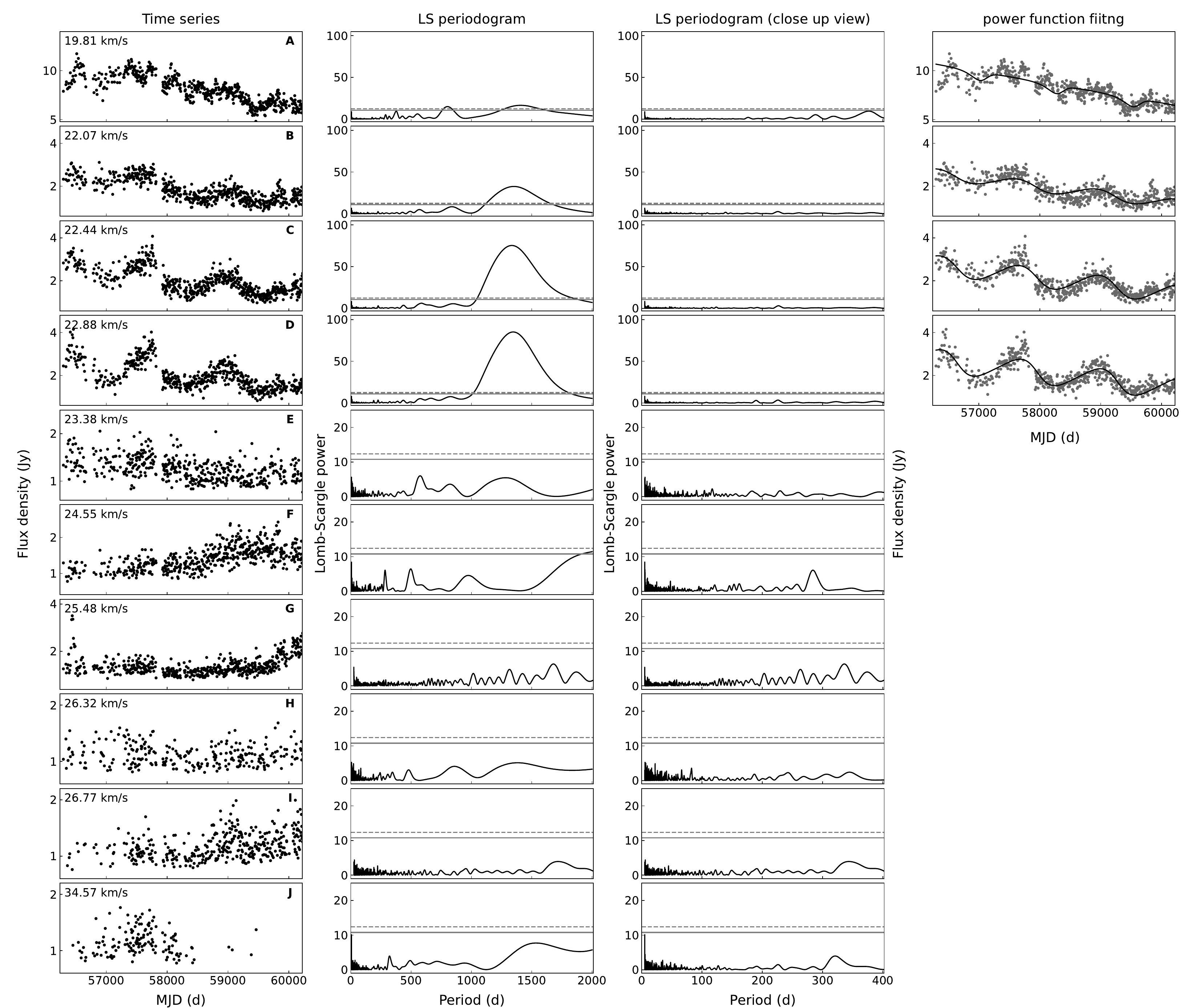}
   \caption{Same as in figure \ref{ana1}, but for G12.209$-$0.102. 
 }
    \label{ana3}
 \end{center}
\end{figure*}

The LS method detected periodicity in four of the 10 features and results of the periodic analysis are summarized in table \ref{tab3} and figure \ref{ana3}.
Continuous and long periodicity is found at all periodic features.
The estimated periods from the LS method for the A, B, C, and D features are 1280, 1280, 1264, and 1264 d, respectively, while those from PF fitting are 1283, 1310, 1285, 1273 d.
The periods of each velocity feature obtained by the two methods are consistent within the estimated errors.
The averaged period with the reliable LS power for each velocity feature is 1272 d.
The results of the fitting show that the flux variation pattern of these features are symmetric with $f = $ 0.40, $-$0.27, 0.31, and 0.26 for A, B, C, and D features, respectively.

\begin{table*}[!h]
\begin{center}
\centering
\caption{Parameters of the periodic features in G12.209$-$0.102}
  \begin{tabular}{cccccc}
  \hline
  label&$V_{\rm LSR}$&$ P_{\rm LS} $&$P_{\rm PF}$&$f$&FWHM\\
  &(km s$^{-1}$)&(d)&(d)&&(d)\\\hline
  A&19.81 &   880\ (58), \textbf{1280}\ (213) & 1283\ (32) &  0.40\ (0.21) &682 \\
  B&22.07 &  878\ (51), \textbf{1280}\ (244) & 1310\ (26) &  0.27\ (0.18) &706 \\
  C&22.44 & 1264\ (216)& 1285\ (13) & 0.31\ (0.09)   &659\\
  D&22.88 & 1264\ (207)& 1273\ (12) & 0.26\ (0.08)   &670
  \\\hline
  \end{tabular}
  \label{tab3}
  \end{center}
\end{table*}

\subsection{G13.657$-$0.599}

\begin{figure}[!h]
 \begin{center}
  \includegraphics[width=8.5cm,bb= 0 0 576 432 ]{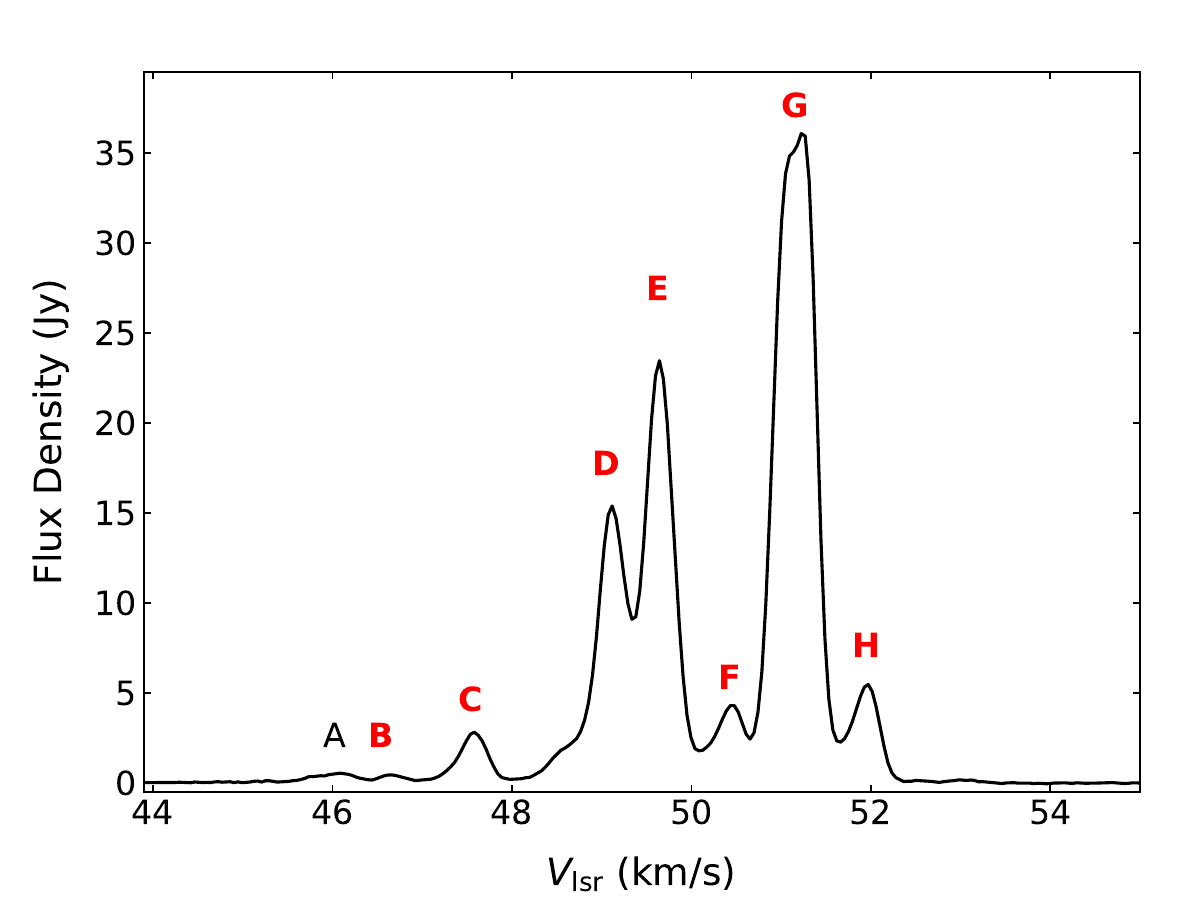}
   \caption{Same as in figure \ref{sp1}, but for G13.657$-$0.599.
   All 601 scans (MJD = 56303--60217) of 5 minutes integration are averaged and 3$\sigma$ detection limit is 0.038 Jy.
   Labels A to H indicate spectral components at $V_{\rm LSR} =$ 45.96, 46.63, 47.58, 49.12, 49.64, 50.45, 51.22, and 51.96 km s$^{-1}$
   , respectively.
   In seven features colored in red (except A), the periodicity is detected in this paper.}
    \label{sp4}
 \end{center}
\end{figure}

\begin{figure*}[h!tb]
 \begin{center}
  \includegraphics[width=17.5cm, bb=0 0 1584 1066]{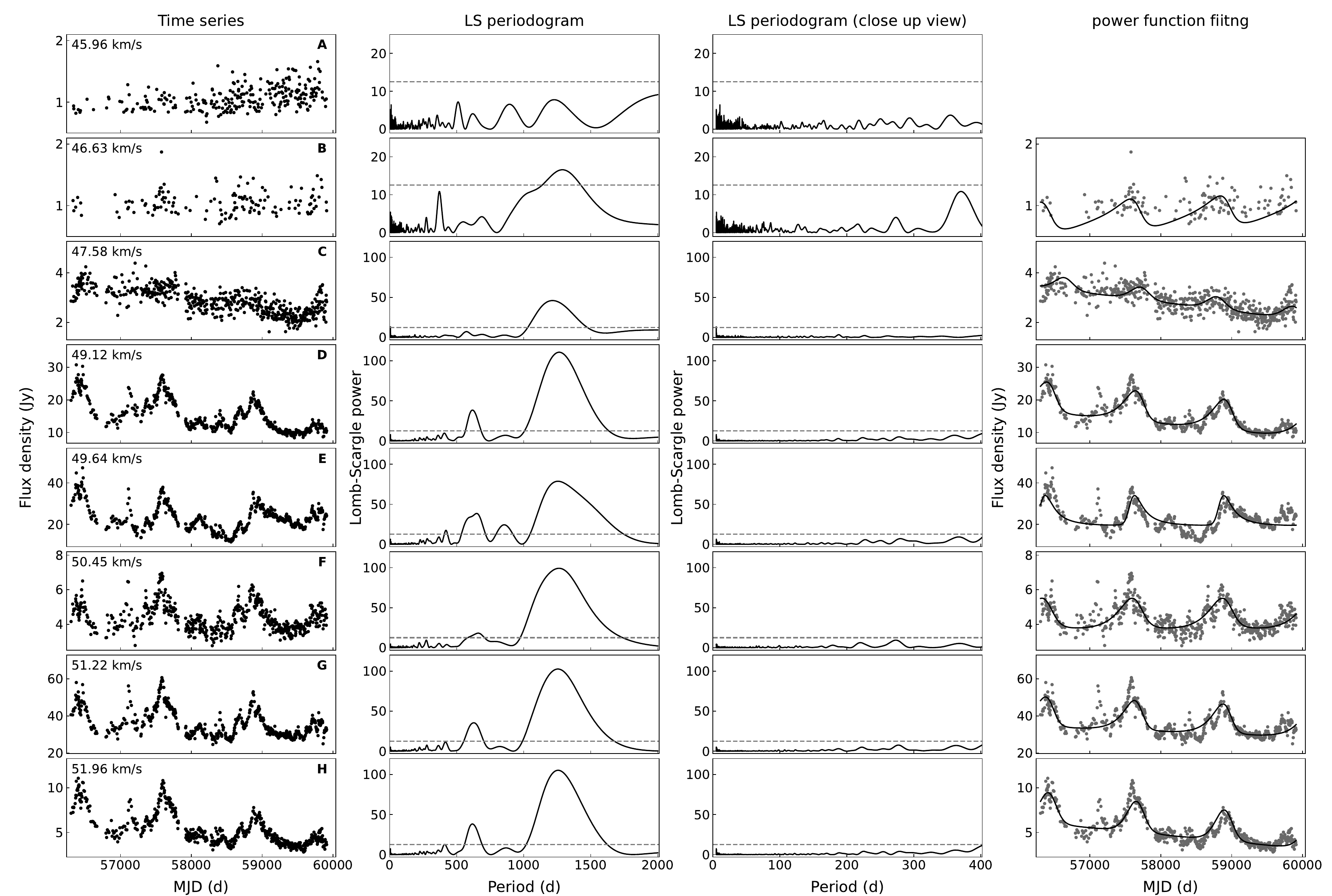}
   \caption{Same as in figure \ref{ana1}, but for G13.657$-$0.599. 
 }
    \label{ana4}
 \end{center}
\end{figure*}

The averaged spectrum obtained by the integration of all 566 scans of G13.657$-$0.599 is presented in figure \ref{sp4}.
In this source, eight velocity features are distributed from $V_{\rm LSR}\ \sim$46 km s$^{-1}$ to $\sim$52 km s$^{-1}$.
Among the eight features in this source, our periodic analysis detected seven periodic features (table \ref{tab4} and figure \ref{ana4}). 
All the velocity features have periods of $\sim$1250 d in this source and the detected periods by two methods are consistent within estimated errors.
The averaged period with the reliable LS power for each velocity feature is 1266 d.
The variation patterns are intermittent except for the features B.
Asymmetric parameters are range from $-$0.67 to 0.66 with symmetric and asymmetric variations.

\begin{table*}[!h]
\begin{center}
\centering
\caption{Parameters of the periodic features in G13.657$-$0.599}
  \begin{tabular}{ccccccc}
  \hline
  label&$V_{\rm LSR}$&$ P_{\rm LS} $&$P_{\rm PF}$&$f$&FWHM\\
  &(km s$^{-1}$)&(d)&(d)&&(d)\\\hline
  \ B&46.63 &   1305\ (215) & 1288\ (21) &  $-$0.67\ (0.13) &648 \\
  \ C&47.58 &   1228\ (187) & 1164\ (15) &  $-$0.35\ (0.14) &353 \\
  \ D&49.12 & 1278\ (182)& 1260\ (4) & $-$0.51\ (0.05)   &344\\
  \ E&49.64 & 1228\ (187)& 1247\ (4) & 0.66\ (0.06)   &295\\
  \ F&50.45 & 1291\ (211)& 1273\ (6) & $-$0.32\ (0.08)   &338\\
  \ G&51.22& 1278\ (176)& 1260\ (5) & $-$0.45\ (0.07)   &304\\
  \ H&51.96 & 1252\ (174)& 1254\ (5) & $-$0.31\ (0.08)   &298
  \\\hline
  \end{tabular}
  \label{tab4}
  \end{center}
\end{table*}

\section{Discussion}

\subsection{Spatial distribution of maser features}\label{mop}
The 6.7 GHz methanol maser emission of our sample was imaged by Hu et al.\ (2016) with the Karl G. Jansky Very Large Array (VLA).
Hu et al.\ (2016) simultaneously observed continuum emission with 4.9840 to 6.0080 GHz and from 6.6245 to 7.6485 GHz and suggested that the H\emissiontype{II} region is associated with all the sources except for G13.657$-$0.599. 

According to Hu et al.\ (2016), 
maser features of G06.795$-$0.257 are distributed in an elliptical morphology.
Sugiyama et al.\ (2016) presented  internal proper motions of maser features in G06.795$-$0.257 using the East-Asian VLBI Network (EAVN).
They concluded that the distribution of the maser features 
are elliptical at the radius of 1260 au on the disk with a position angle of the semi-major axis of $-$140$\degree$ and an inclination of 60$\degree$, and the motions of maser features point counterclockwise rotation and expansion along the elliptical.

The spatial distribution of maser features in  G10.472$+$0.027 is localized at 3 positions.
There are two blue-shifted ($V_{\rm LSR}\sim$60 km s$^{-1}$) components located at 0.3 arcsec east and at 0.5 arcsec west from the red-shifted components.
The size of spatial distribution is 0.8 arcsec $\times$ 0.2 arcsec, which correspond to 6720 au $\times$ 1680 au when the distance of 8.4 kpc is adopted.
The blue-shifted features with a period of $\sim$1600 d are found at both east and west components.

The maser features in G12.209$-$0.102 are distributed in arc-like with a diameter of 2.6 arcsec, which correspond to $\sim$35000 au when the distance of 13.4 kpc is adopted.
This value is an order of magnitude larger than the typical size of distribution of masers associated with a single source (Bartkiewicz et al.\ 2016).
Therefore, these maser features may be associated with multiple sources.
The 20 to 23 km s$^{-1}$ features for which the period was detected in this study are found in all different positions of the arc.

The maser features in G13.657$-$0.599 are distributed in a linear shape along with the north-west to south-east direction.
The size of spatial distribution is 0.3 arcsec $\times$\ 0.4 arcsec, which correspond to 1350 au $\times$ 1800 au when the distance of 4.5 kpc is adopted.
The velocity field is complex and no gradients are seen in this source.

In G10.472$+$0.027 and G12.209$-$0.102,  it is difficult to compare the spatial distribution of maser spots with  spectrum taken by our single-dish monitoring, because there are multiple components with the same velocity but in different locations.
VLBI monitoring is indispensable to understand which maser components are variable.

\subsection{Maser periodicity}\label{pe}
We detected new maser periodicity in four HMSFRs.
All sources have at least two features with periods longer than 1000 d.
According to TYM23 and references therein,
among the 29 known periodic maser sources, the longest period is $\sim$1260 d for G05.900$-$0.430, and this is the only source with a periodicity longer than 1000 d. 
The variation patterns of the features with long periodicity are continuous and symmetric in G10.472$+$0.027 and G12.209$+$0.102, and intermittent in G06.795$-$0.257 and G13.657$-$0.599.
 G10.472$+$0.027 has three periods that are not in harmonics.
 The cases of multiple periods in one source have been reported in G9.621$+$0.20E (MacLeod et al.\ 2022) and G05.900$-$0.430 (TYM23).
There are several explanations of the mechanism for such multiple periods, such as Dicke's superradience proposed by Rajabi et al.\ (2023), but the details have not been clarified.
In the following subsections, we focus only on the periods of about 1000 d or more.

\begin{table*}[Htb]
\begin{center}
\caption{Property of sources}
  \begin{tabular}{cccccccccc}
  \hline
Source&$P$\footnotemark[*]&$n_{\rm P}/n$\footnotemark[$\dag$] & Variation\footnotemark[$\ddag$]& Luminosity\footnotemark[$\S$]&Distance&Dist\footnotemark[$\|$]&H\emissiontype{II} region\footnotemark[$\#$]&morphology&size\\
  & (d) & &pattern&${\rm log}(L/L_{\odot})$& (kpc)&\ ref&&&(au)\\\hline\
  G06.795$-$0.257 & 968 &13/15&I& 4.0  & 3.0 &A&Y&elliptical& 2520\footnotemark[$**$]\\
  G10.472$+$0.027 & 1624 $\dag\dag$&11/13\footnotemark[$\ddag\ddag$]&C&5.7 &8.6&B&Y&3 groups&6720 × 1680\\
  G12.209$-$0.102 & 1272 &4/10&C&5.5 &13.4&A&Y&arc-like&35000\\
  G13.657$-$0.599 & 1266 &7/9&I\footnotemark[$\S\S$]& 4.4 &4.5 &A&N&linear&1350 × 1800\\
  \hline
  \end{tabular}
  \label{statab}
  \end{center}
   \begin{tabnote}
  \footnotemark[$*$]The averaged periods obtained by the LS method in each source.\\
  \footnotemark[$\dag$] Numbers of periodic spectral features/numbers of all spectral features.\\
  \footnotemark[$\ddag$]`I' and `C' represent intermittent variation and  continuous  variation, respectively.\\
  \footnotemark[$\S$]Urquhart et al.\ (2018)\\
  \footnotemark[$\|$]Reference for adopted distance. A: Reid et al.\ (2016) estimated the shapes of the spiral arm by the Bayesian approach. B: Sanna et al.\ (2014) estimated by trigonometric parallax.\\ 
  \footnotemark[$\#$]Hu et al.\ (2016)\\
  \footnotemark[$**$]Diameter when reprojected onto a circle (Sugiyama et al.\ 2016).\\
\footnotemark[$\dag\dag$]The  averaged periods of four features with a period of over 1500 d.\\
  \footnotemark[$\ddag\ddag$] Four spectral features have the periods of over 1500 d, another three have the periods of 366 d, and the other four have the periods of $\sim$1000 d.\\
  \footnotemark[$\S\S$]One feature has the continuous variation pattern.
  \end{tabnote}
\end{table*}

\subsubsection{Protostellar pulsation instability}

\begin{figure}[htb]
\centering
    \includegraphics[width=8cm, bb =  0 0 576 360]{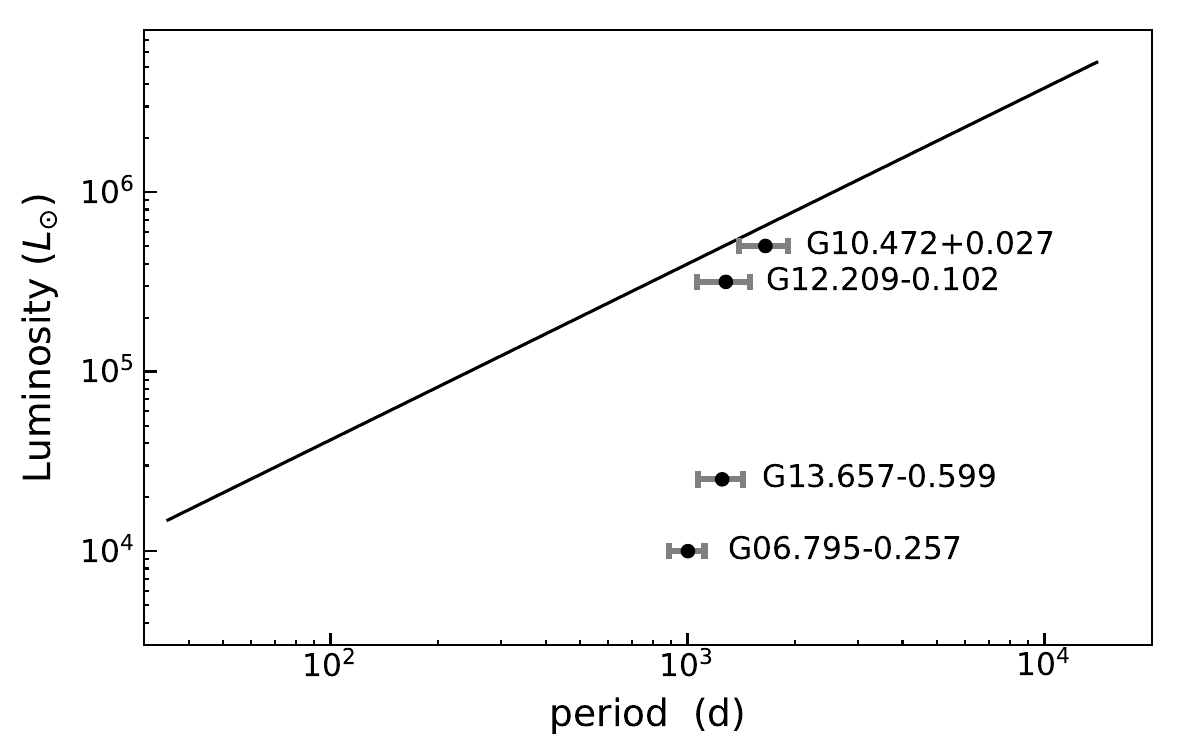}
    \caption{Protostellar luminosity versus estimated maser periods.
     The solid line represents the PL relation given by Inayoshi et al.\ (2013).
     According to Urquhart et al.\ (2018), the total uncertainty of luminosity due to measurement error and SED fitting is $\sim$100\%, which does not include uncertainty of the distance of the sources.}
    \label{PL}
\end{figure}

Inayoshi et al.\ (2013) proposed that the high-mass protostars with large mass accretion with rates of $\dot{M_{*}}\gtrsim 10^{-3}\ M_{\odot}\ {\rm yr}^{-1}$ become pulsationally unstable.
This instability is caused by the $\kappa$ mechanism in the He$^+$ layer, where the radiative energy flux is blocked and converted into the pulsation energy (e.g., Cox 1980; Unno et al. 1989; Inayoshi et al. 2013).
During this pulsation phase, the luminosity of the protostars varies periodically and the temperature of the surrounding material rises to and falls from a temperature suitable for maser pumping, resulting in that the maser fluxes increase and decrease.
In this model, the flux variation pattern of the associated maser is expected to be continuous, thus the protostellar pulsation instability model may well describe the variation in G10.472$+$0.027 and G12.209$-$0.102.
Inayoshi et al.\ (2013) also derived the period-luminosity (PL) relation 
\begin{equation}
{\rm log}\frac{L}{\ L_{\odot}} = 4.62 + 0.98\ {\rm log}\frac{P}{100 \ {\rm days}},
\end{equation}
where {\it L} is luminosity of the protostar and {\it P} is period expected from the maser variation.
Figure \ref{PL} plots the estimated periods versus protostellar luminosity for each source (see also Table \ref{statab}).  
The solid line represents the PL relation given by equation (3).
In figure \ref{PL}, the central protostar luminosity and periods of maser flux variation are roughly consistent with the expected PL relation for G10.472$+$0.027 and G12.209$-$0.102 which suggests that the mechanism of maser flux variability of these two sources can be explained by protostellar pulsation instability.

If the pulsationally unstable model is applied for these sources, the protostellar mass and mass accretion rate estimated from the detected period using the equations (2) and (4) in Inayoshi et al.\ (2013) are $M_{*}=41\ M_{\odot}$ and $\dot{M_{*}}=2.4\times$10$^{-2}\ M_{\odot}\ {\rm yr}^{-1}$ for G10.472$+$0.027 and $M_{*}=38\ M_{\odot}$ and $\dot{M_{*}}=2.0\times$10$^{-2}\ M_{\odot}\ {\rm yr}^{-1}$ for G12.209$-$0.102, respectively.

On the other hand, the sources with intermittent variable pattern (G06.795$-$0.257 and G13.657$-$0.599) have an order of magnitude smaller luminosity than those expected from PL relation.
Thus variation mechanisms of these sources may be originated from binary system.

\subsubsection{CWB}

In the CWB, modeled by van der Walt\ (2011), periodic wind interaction in a binary system generates changes in the background free-free emission which is amplified by the masers.
It requires a varying H\emissiontype{II} region.
The free-free radiation acts as the seed photons.
Since our sample except for G13.657$-$0.599 are associated with the H\emissiontype{II} region (see table \ref{statab}), CWB model is a candidate of the mechanism of the flux variation of G06.795$-$0.257.
According to van der Walt (2011) and van der Walt et al.\ (2016), the maser variation profile explained by the CWB model may be intermittent with long decaying time.
Because of the intermittent variation pattern, CWB model seems appropriate for G06.795$-$0.257, however, in this model, only photons moving towards us along the line of sight contribute to the variation in maser flux.
Therefore, the maser features located beyond the central protostar from our point of view may not be variable.
In G06.795$-$0.257, according to Sugiyama et al.\ (2016), each maser features are distributed concentrically in circumstellar disk and the red-shifted features which range from 24 km s$^{-1}$ to 31 km s$^{-1}$ are in the far side of the disk.
Since the red-shifted features (H to O) on the far side of the disk have a periodicity, we can conclude that the CWB model is not appropriate for G06.795$-$0.257.

\subsubsection{Eclipsing binary}
Maswanganye et al.\ (2015) proposed that maser flares in G358.46$-$0.391 and G338.93$-$0.06 are due to variations in the free$-$free background seed photon flux caused by an eclipsing binary.
As in the case of CWB, this model requires an H\emissiontype{II} region that produces seed photons, and only the maser features on the front side of the H\emissiontype{II} region causes flare.
Therefore, eclipsing binary model cannot explain the maser flux variation in G06.795$-$0.257.
G13.657$-$0.599 is not associated with any H\emissiontype{II} regions, thus eclipsing binary model cannot appropriate for this source.

\subsubsection{Spiral shock}
Rotating spiral shock model in a binary system is proposed by Parfenov and Sobolev (2014).
The dust temperature variations are caused by rotation of hot and dense material in the spiral shock wave in the central gap of circumbinary disk.
This model require neither infrared emission variability nor radio continuum, but needs an edge-on protostellar disk in which masers are associated.
In G06.795$-$0.257, as mentioned in the previous subsection, the maser features are distributed in an elliptical morphology.
On the other hand, the spatial distributions of the maser features associated with  G13.657$-$0.599 is linear, but the velocity gradient is clearly different from that of an edge-on disk.
Therefore spiral shock model can not appropriate for G06.795$-$0.257 and G13.657$-$0.599.

\subsubsection{Periodic accretion}
Interaction between a binary system and accreting matter can strongly modulate accretion on to the central protostar (e.g., Artymowicz \& Lubow 1996; Mu\~noz \& Lai 2016).
Araya et al.\ (2010) proposed that such modulated accretion causes the periodic variability in some methanol maser sources through variations of the dust temperature.
Szymczak et al.\ (2016) and  Olech et al.\ (2019) suggested that this process is responsible for the periodic variability of G107.298$+$5.639 and G59.633$-$0.192.
In this periodic accretion model, material from the circumbinary disk is accreted onto the protostars, heating the dust and increasing the infrared radiation field (Araya et al.\ 2010), thus a maser can be amplified anywhere in the disk regardless of its geometry.
Therefore, periodic accretion model may be appropriate for G06.795$-$0.257 and G13.657$-$0.599, where almost all the features are periodic with same periods.

We summarized the results of the discussions in the subsections from 4.2.1 to 4.2.5 in table \ref{sumtab}.
In addition to our single-dish monitoring observations, monitoring observations with Very Long Baseline Interferometry  (VLBI) to obtain the flux variations for each spatial component will lead to a better understanding the nature of each source.

\begin{table*}[htb]
\begin{center}
\caption{Variation mechanism for each source}
  \begin{tabular}{cccccc}
  \hline
Source&Pulsation&CWB&Eclipse binary&Spiral shock&Periodic accretion\\\hline
  G06.795$-$0.257 & $\times$  & $\times$  &$\times$&$\times$&$\circ$\\
  G10.472$+$0.027 & $\circ$  &-&-&-&-\\
  G12.209$-$0.102 & $\circ$ &- &-&-&-\\
  G13.657$-$0.599 & $\times$ & $\times$ &$\times$&$\times$&$\circ$ \\
  \hline
  \end{tabular}
  \label{sumtab}
  \end{center}
   \begin{tabnote}
  `-' Not evaluated in this paper.
  \end{tabnote}
\end{table*}

\section{Summary}
We have presented the results of the 10-years monitoring of four periodic 6.7 GHz Class II methanol maser sources.
The detected periods in our sample range from 968 to 1624 d, although the detected period is tentative due to the short monitoring term relative to the estimated period.
While the cause of the periodicity is yet to be confirmed, it seems likely that G10.472$+$0.027 and G12.209$-$0.102 may be explained by protostellar pulsation instability since their flux variable profiles show symmetric sine curves and that the central protostar luminosity and periods of maser flux variation are consistent with the expected PL relation.
G06.795$-$0.257 and G13.657$-$0.599 have luminosity  smaller than those expected from PL relation,
suggesting a variation mechanism of these sources from binary system.
Since almost all the velocity features obtained by the single-dish observations vary with the same period, it is possible that the periodic variations of these sources can be explained by a periodic accretion model.
VLBI monitoring observations with high spatial resolution will lead to a better understanding of the nature of these sources.

\begin{ack}
The authors are grateful to all the postdocs and students at Ibaraki University who have supported observations of the Ibaraki 6.7 GHz Methanol Maser Monitor (iMet) program.
This work is partially supported by the Inter-university collaborative project ``Japanese VLBI Network (JVN)'' of NAOJ. This study benefited from financial support from the Japan Society for the Promotion of Science (JSPS) KAKENHI program (21H01120 and 21H00032).

\end{ack}

\appendix

\section*{Phase-folded light curves}
\begin{figure*}[h!tb]
 \begin{center}
  \includegraphics[width=17.5cm, bb=0 0 2610 2412]{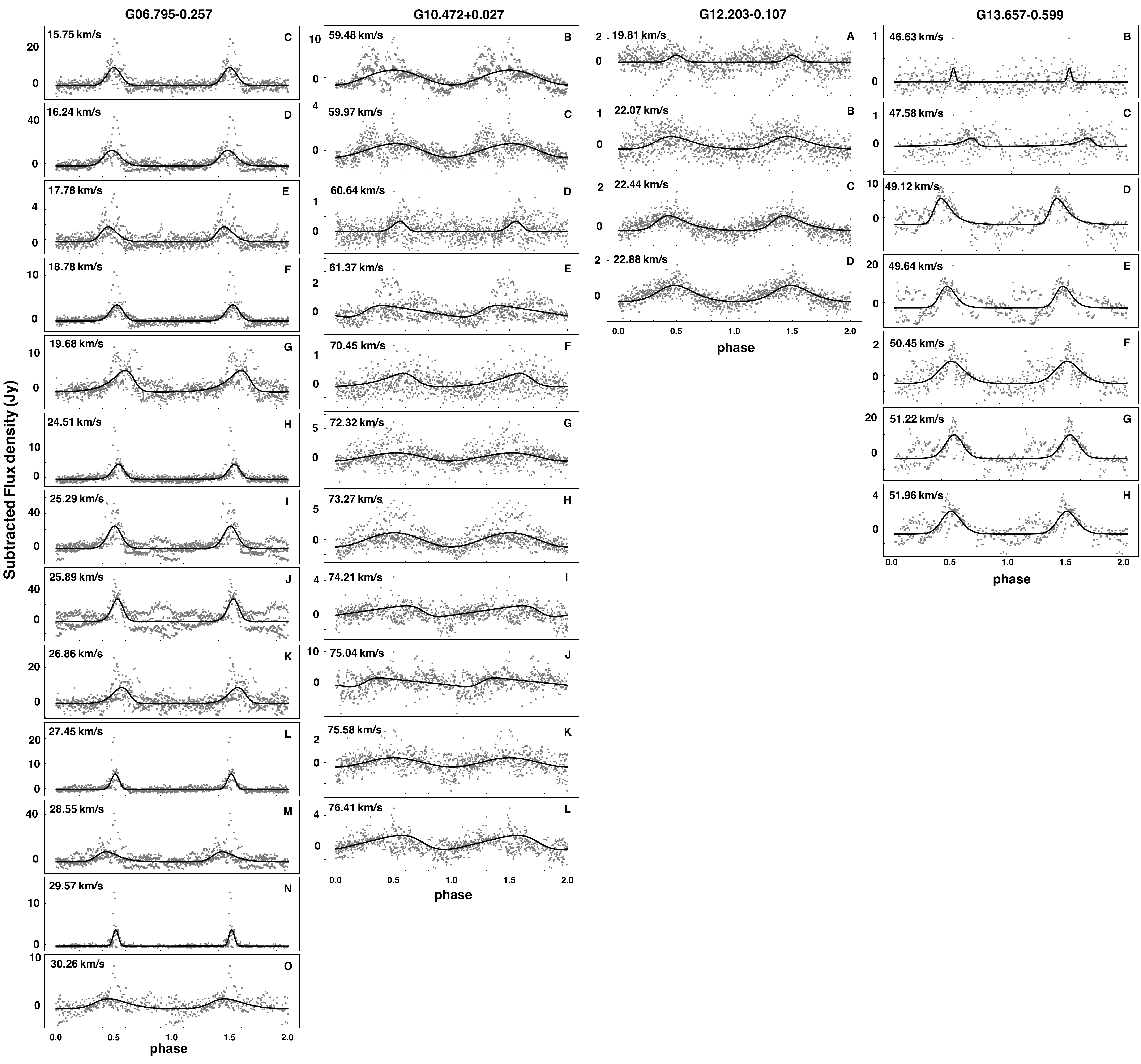}
   \caption{Phase-folded and trend-subtracted light curves for each source. 
    The solid curves represent the best fits of asymmetric power function to the data.
 }
    \label{phase}
 \end{center}
\end{figure*}

We show in this appendix, the phase-folded and trend-subtracted light curves of all the periodic features (Figure \ref{phase}).
Here, phase-folded light curve is created by subtracting the components that can be fitted by a linear function. 
Therefore, the average value of the subtracted flux density is $\sim$0.
The solid curves represent the best fits of a asymmetric power function to the data.
The results show a tendency that these are various patterns in one source, similar to the fitting to the light curve for the entire data.

{}

\end{document}